\documentclass[journal,11pt,onecolumn,draftclsnofoot]{IEEEtran}
\def\BibTeX{{\rm B\kern-.05em{\sc i\kern-.025em b}\kern-.08em T\kern-.1667em\lower.7ex\hbox{E}\kern-.125emX}}
\usepackage{setspace} 
\usepackage{times}
\usepackage{gensymb}
\usepackage{tabularx}
\usepackage{lipsum}
\usepackage{array}
\usepackage{booktabs}
\usepackage{listings}
 \usepackage{float}
 \usepackage{xcolor}

\usepackage{enumitem}

\usepackage{multirow}

\ifCLASSINFOpdf
 \usepackage[pdftex]{graphicx}
 \else
 \usepackage[dvips]{graphicx}

 \usepackage{amssymb,amsmath}
 
\fi

\usepackage[caption]{subfig}
\usepackage{cite}

\usepackage[export]{adjustbox}

\lstdefinestyle{lststyle}{
 captionpos=b, 
 tabsize=2,
 basicstyle=\linespread{0.9}\footnotesize\ttfamily,
}
  
\begin{document}

\noindent

\title{Internet-of-Things Devices and Assistive Technologies for Healthcare: Applications, Challenges, and Opportunities}

\author{\IEEEauthorblockN{Marc Jayson Baucas, Petros Spachos, and Stefano Gregori}

\IEEEauthorblockA{School of Engineering, University of Guelph, Guelph, Ontario, Canada} 
}

\maketitle
\begin{abstract}
\noindent Medical conditions and cases are growing at a rapid pace, where physical space is starting to be constrained. Hospitals and clinics no longer have the ability to accommodate large numbers of incoming patients.  It is clear that the current state of the health industry needs to improve its valuable and limited resources. The evolution of the Internet of Things (IoT) devices along with assistive technologies can alleviate the problem in healthcare, by being a convenient and easy means of accessing healthcare services wirelessly. There is a plethora of IoT devices and potential applications that can take advantage of the unique characteristics that these technologies can offer. However, at the same time, these services pose novel challenges that need to be properly addressed. In this article, we review some popular categories of IoT-based applications for healthcare along with their devices. Then, we describe the challenges and discuss how research can properly address the open issues and improve the already existing implementations in healthcare. Further possible solutions are also discussed to show their potential in being viable solutions for future healthcare applications. 
\end{abstract}

\section{Introduction}
\noindent The recent pandemic has revealed underlying limitations, inequities, and gaps in universal healthcare access. COVID-19 outbreak has exposed the centrality of health, both as an outcome and as an engine for economic and social development. Even well-developed healthcare systems face unprecedented challenges due to demographic, epidemiological, and health transitions. The hard-learned lessons for the health sector dictate improvements in both the effectiveness and efficiency of services. Healthcare services need to take advantage of new technologies that can upgrade the capacity to prevent, diagnose, and treat diseases. A promising concept that is rapidly spreading in the healthcare sector due to its ability to effectively integrate its services is the Internet of Things (IoT)~\cite{hc-iot}. IoT enables people and objects to be connected anytime, anyplace, anywhere, via any network and service. IoT networks can operate within healthcare centers, they can be scaled up to provide services to urban communities, and they can even connect patients and healthcare providers over longer distances, thereby lowering geographical barriers to healthcare access and the burden on healthcare facilities.

E-health services relying on electronic devices and smart environments~\cite{hassan} are leading to the proliferation of assistive technologies based on wireless networks and IoT implementations~\cite{hc-iot}. These technologies equip patients with devices that monitor and aid certain aspects of their health and well-being. The collected data is shared between an IoT device and a central server, usually through a wireless network. Data hierarchy is constructed to allow patients to be remotely screened for any anomalies and aided for certain conditions concerning their health.  

Although IoT devices along with assistive technologies provide several benefits to healthcare services, they also pose challenges, specifically in patient privacy, data latency, service interactability, device constraints, and scalability. The privacy and security of a patient's personal information have always been a concern in healthcare services~\cite{shuai}. IoT-based services make use of wireless communication to transfer data. This type of communication has its vulnerabilities, which may result in user information being compromised~\cite{Abbas}. Aside from privacy, wireless communications have a bottleneck in terms of data transmission. Continuous transmission of data could result in the exhaustion of the server and in the congestion of the network~\cite{rpm-iot-ecg}. Also, for applications that need data to reach the server in real-time, a discontinuous dataflow could be detrimental to the service. As for interactability, some assistive technologies are passive and only collect or transmit data. The health center is then responsible for translating this data into useful information to screen or diagnose the patient. However, the latency due to data transmission and processing prevents health centers to respond in real-time, making it hard to detect and respond to emergencies. A system that allows real-time interactions between the health center and the patient can allow better, more responsive, and reliable services. 

At the same time, the constraints of the devices used in collecting and transmitting the data limit the effectiveness of the service, in particular with regard to the ability of the devices to keep up with the demands of the service in terms of power consumption and processing requirements over time. Finally, the scalability of the overall service is also important as it defines its capacity in terms of how many users it can handle. An application that cannot scale along with the growth of its devices and users will not be able to keep up with the need of the communities that use it. As a result, services that suffer from device constraints and scalability issues lose their credibility and viability as products in the healthcare industry. 

This article reviews the available IoT-based services in healthcare and discusses some representative IoT-based healthcare applications along with their challenges. Then, the research opportunities that address these issues are discussed. As such, this article will cover the following four main areas:
\begin{enumerate}[label=(\roman*)]
\item  Overview of IoT-based Applications in Healthcare, where we present some popular IoT-based applications for healthcare, along with their advantages.  Then,  we discuss the main domain of each application and we further discussed three different approaches for their data processing location.
\item Devices and Assistive Technologies in IoT-based healthcare, where we discuss IoT devices and assistive technologies and how they can enhance the current state of the art in health care. We present three main types along with examples.
\item Challenges in IoT-based Healthcare Services, where we discuss five important challenges of IoT-based health care services.
\item Research Approaches and Opportunities, where we focus on existing solutions for the challenges and we further discuss potential opportunities for the researchers.

\end{enumerate}

\section{Overview of IoT-based Applications in Healthcare}\label{iot}
\noindent
 
\begin{table}[t!]
    \centering
    \normalsize 
    \begin{tabular}{|p{8cm}|p{8cm}|}
        \hline
        \textbf{IoT-based Applications in Healthcare} & \textbf{Advantages} \\ \hline
         Vital signs monitoring through mobile phone~\cite{sood} & Wide area and scope of transmission, optimized operating system for managing data \\ \hline
          Electronic health records reading with smart devices~\cite{rpm-auto} & Portable, easy-to-learn user interface, convenient for users who already own smart devices  \\ \hline
         Wearable electrocardiogram (ECG)~\cite{work-low, rpm-iot-ecg} & Low costs, compact, optimized power usage \\ \hline
         Robots and drones for patient assistance~\cite{miseikis} & Easy to train, precise movements and actions, make repetitive tasks convenient, automative  \\ \hline
         Application specific wireless sensors for emergency detection (i.e. fall prediction, pain detection)~\cite{saadeh, yang-pain}  & Wearable, modular, can be designed to fit patient \\ \hline
    \end{tabular}
    \caption{IoT-based applications in healthcare along with their advantages.}
    \label{ex_ad}
\end{table} 

\noindent
Healthcare is an industry that adopted IoT to extend its scope and expand medical services into the wireless era~\cite{hc-iot}. The results are IoT-based healthcare applications, where a patient can be monitored, treated, and aided by a medical center with the use of IoT devices over the network. These applications provide many advantages towards the improvement of healthcare services. Some IoT-based applications in healthcare along with their advantages are shown in Table~\ref{ex_ad}.

\subsection{IoT Use Cases in Healthcare}

\subsubsection{Remote Patient Monitoring}
Remote Patient Monitoring (RPM) is a type of remote telehealth service that uses medical devices to observe and treat patients regardless of distance. RPM services reduce the need for patients to travel to diagnose illnesses or undergo check-ups since the wireless network can achieve both~\cite{hassan}. By combining the sensors through the wireless network, it creates a wireless sensor network (WSN), which is a type of IoT system, that makes use of sensors that are programmed to collect data~\cite{work-trans}. In this way, it creates a sensing network for a diverse array of raw data. Healthcare services use this design to create an RPM service using the data collected from its wireless sensors. This approach complements the RPM by improving the quality of its service. The simplicity of the sensors used in the network yields faster transmission rates of collected data. As a result, the service facilitates real-time sensing and monitoring. The use of more optimized sensors to allow lesser energy consumption and better operation times suggested by~\cite{rpm-iot-ecg}. By using simpler and application-specific sensors, RPM services can be more energy and resource-efficient. Their work shows the advantages of using a compressed ECG sensor to collect and deliver real-time data. This design results in an envisioned RPM service having a longer battery life due to its efficient design and lesser processing demands.

 Another example of a proposed framework that uses WSNs as its RPM system is in~\cite{rpm-comm}. They create a community-based WSN using LoRaWAN to increase the coverage of the RPM service. With it, they can conduct medical screenings for conditions, specifically urinary tract infection (UTI). RPMs already benefit from being able to monitor patients over long distances. However, due to the wide-area coverage of LoRaWAN and the simplicity of the WSN, they create a service that can span their whole community. This design shows the benefits of scalability in IoT in healthcare and how it improves the network's overall scope.
 
\subsubsection{Smart Hospitals} 

A smart hospital is a type of IoT use case in healthcare where medical devices connect through a wireless network to improve and enhance their services. As a result, it creates an ecosystem where these devices can carry out tasks and operations related to e-health under a healthcare server. As their services incorporate IoT systems, smart hospitals can improve their efficiency and lower healthcare cost, as in~\cite{Catarinucci}. With this design, monitoring and treating patients become more self-sustaining since the system can automate the tasks of monitoring and treating patients. 

A proposed solution for reducing the risks of the COVID-19 outbreak by using smart hospitals is in~\cite{jaishwal-covid}. They suggest using smart technology and its properties reduce the risk of face-to-face interaction and physical contact. Instead of physically monitoring their patients, a system with wireless capabilities can make use of these devices. These services also benefit from being able to automate the monitoring process in medical treatment. A discussion of using smart hospitals to improve the quality of service that treats and cares for the elderly is in~\cite{maresova}. A high percentage of hospital patients are elderly who suffer from conditions that impair their movement, which makes travelling more challenging. They suggest the use of smart technology to bring the treatment to the patient. This usage results in a patient's place becoming a part of the virtual hospital, and the clinic is now able to administer adequate care to the user. Through smart computing and automation, medical centers can craft each treatment to cater to different diseases and conditions.

\subsubsection{Mobile E-health} 

Mobile e-health is an IoT use case for healthcare that focuses more on the wireless medium that connects mobile devices for medical services~\cite{ahad-5g}. Its architecture has stationary gateways that are fixed transceivers called cell towers. Each tower is a large antenna that is strategically placed overland to expand the network. Healthcare services make use of this technology to extend their services over long distances. A health monitoring system that makes use of mobile computing is in~\cite{rpm-auto}. It takes advantage of the large distances covered by a mobile network to connect patients to their healthcare services. Mobile e-health allows medical services to monitor and communicate with their patients through their phones. It is beneficial in areas that only cellular networks can reach. Therefore, medical centers can reach their patients, especially during emergencies. Aside from its network of mobile devices, its design incorporates wireless sensors. Therefore, their design highlights the interchangeability of the parts of each IoT use case. These systems have no hard constraints but are flexible architectures. This design opens the medical field to create combinations that can improve the quality of the healthcare service. 

Another example makes use of a 5G-based mobile network to expand its remote healthcare services~\cite{ahad-5g}. This design combines the mobile network with smart devices. It incorporates the automation capabilities of a smart hospital and the distance coverage of mobile e-health. Also, it highlights the advantages of each use case and IoT as a whole when expanding remote healthcare and its services.

These implementations of IoT in healthcare are summarized in Table~\ref{iot-table}, where we highlight the domain, target IoT devices, and integrated IoT systems.

\begin{table}[t!]
    \centering
    \normalsize
    \begin{tabular}{|p{7cm}|p{5cm}|p{3.7cm}|}
        \hline
          \textbf{Domain} & \textbf{Target IoT Device(s)} & \textbf{IoT System(s)}  \\ \hline
         Smart healthcare infrastructure~\cite{Catarinucci} & Biomedical systems & RPM, Smart hospital \\ \hline
        Remote healthcare~\cite{jaishwal-covid}  & Contact-free technologies, Drones, Robots & Smart hospital\\ \hline
         Smart healthcare for the elderly~\cite{maresova}& Smart homes, Smart furniture & Smart hospital\\ \hline
        Real-Time RPM~\cite{work-trans}& Wireless Transducer, ECG & RPM \\ \hline
         Remote ECG Monitoring, Compressive sensing~\cite{rpm-iot-ecg}& ECG & RPM \\ \hline
         Community-wide UTI screening system, Personalized wireless healthcare~\cite{rpm-comm} & Android phone & RPM, Mobile e-health\\ \hline
         Mobile smart healthcare using 5G networks~\cite{ahad-5g}& Smartphones & Smart hospital, Mobile e-health\\ \hline
         Smart neonatal health monitoring system~\cite{rpm-auto}  & Vital sensors, Smartphone & RPM, Smart hospital, Mobile e-health \\ \hline
    \end{tabular}
    \caption{Implementations of IoT systems in healthcare.}
    \label{iot-table}
\end{table}

\subsection{Data Processing}
\noindent

An IoT network is a collection of devices with computing capabilities that share and exchange data within a wireless medium. In combination with a hierarchy of operations and processes, the result is a diverse set of online services. A standard network is composed of a server, a gateway, and a collection of IoT devices~\cite{fog-cloud}. The server is the main control hub of the network, where most of the data is stored. The gateway is the data router of the network that serves as the bridge between the end devices and the cloud server. The end devices, or IoT devices, are the peripherals at the end of the network. In healthcare, these IoT devices are used to collect, transmit, and report medical data depending on their design and purpose.

For data processing, there are three approaches: cloud computing, fog computing, and edge computing. The primary difference is the location where data processing occurs.

\begin{enumerate}[wide, labelwidth=!, labelindent=0pt]
    \item \textit{Cloud-based}: A cloud-based IoT network concentrates all of its resources in the cloud server~\cite{yang-pain}. Most data processing is carried out in this device. The end devices are only used to collect data, which is then sent to the cloud. This centralized hierarchy allows controlling and regulating the service more conveniently. Also, cloud servers have an advantage in resource management due to their single control unit. Since cloud servers are considered the main online data storage of the service, data is more readily accessible from the centralized repository.
    \item \textit{Fog-based}: A fog-based IoT network is used to improve the network traffic by adding fog devices~\cite{sood}. These fog devices function as local servers closer to the source of data. They allow the network to reallocate data packets and redirect data paths to reduce the strain going to the main cloud server. The fog device takes the place of the gateway in a standard IoT setup~\cite{fog-rpm}. It also extends the network in such a way that a new area of operation is added to the hierarchy thereby allowing processes to be moved across the network. 
    \item \textit{Edge-based}: An edge-based IoT network focuses on reallocating processing requirements to devices at the edge of the network~\cite{deng-delay}. This feature gives the IoT devices that collect the data the ability to process it as well. By doing so, analysis is readily provided to the user without the need for continuous online connection to the network~\cite{fog-rpm}. Unlike the other two configurations, edge-based networks can still effectively function offline. It reduces the reliance of the IoT devices towards the main server. However, in return, this configuration requires more processing power from its IoT devices.
\end{enumerate}

\begin{figure}[t!]
	\newcommand{\fscale}{0.28}%
    \centering
	\subfloat[Cloud-based configuration.]
	{   \includegraphics[scale=\fscale,valign=c]{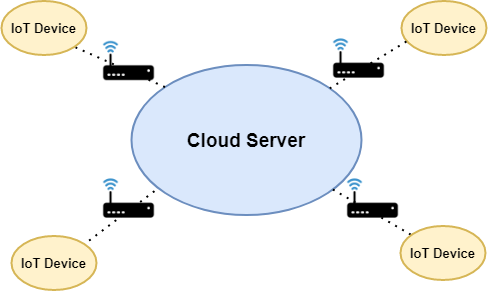}\label{cloudconf}
		\vphantom{\includegraphics[scale=\fscale,valign=c]{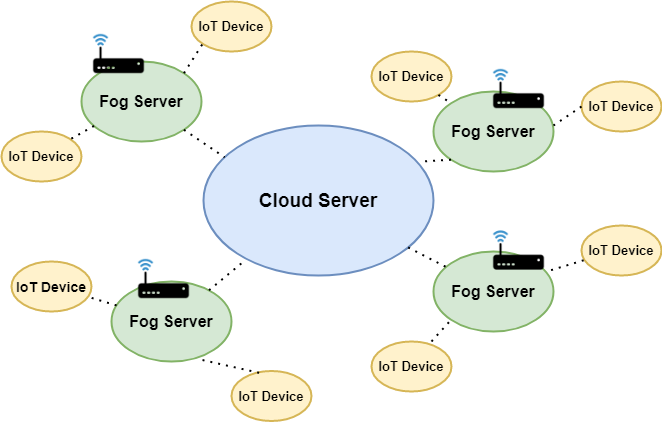}}}
	\subfloat[Fog-based configuration.]
	{    \includegraphics[scale=\fscale,valign=c]{CloudFogB.png}\label{fogconf}}
	\subfloat[Edge-based configuration.]
	{    \includegraphics[scale=\fscale,valign=c]{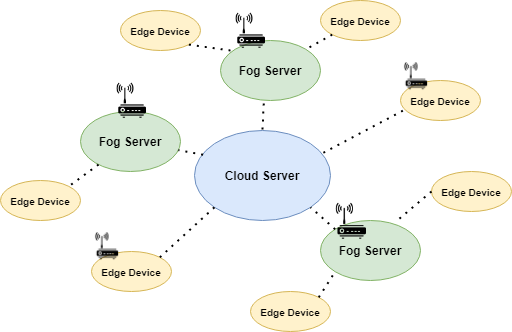}\label{fogconf}
	\vphantom{\includegraphics[scale=\fscale,valign=c]{CloudFogA.png}}}
    \caption{Available data processing approaches.}
    \label{arch}
\end{figure}

The three approaches are shown in Fig.~\ref{arch}.

\section{Devices and Assistive Technologies in IoT-Based Healthcare}\label{app}
\noindent 
Assistive technologies are any devices, software, or equipment  that are used to aid its users in carrying out challenging tasks. In healthcare, they can further categorized depending on their architecture and main purpose~\cite{Catarinucci}. The following describes examples of assistive technologies that are used in IoT-based healthcare.

\subsection{Wearable Devices}
\noindent
Wearable devices are the simpler forms of assistive technology. Their purpose is to collect the designated data that it is programmed to~\cite{rpm-iot-ecg}. Equipped with wireless capabilities to enable RPM services, patients have the option to wear monitoring devices that transmit their vitals to a configured healthcare center. For instance, a wearable ECG sensor allows simultaneous monitoring of multiple patients using the same IoT network~\cite{work-low}. Due to its low-power consumption, it has long battery life. This feature improves the overall quality of service and reliability of the system. Also, the prototype sensor is cost-effective due to the simplicity of the wearable device. 

This design shows the benefits of using this simple form of assistive technology. It can improve the speed of data transmitted to the server by cutting down on the preprocessing. Also, with only one parameter being monitored and do not require patient-server interactions, it can be a cost-effective option for services.
    
\subsection{Mobility and Sensory Aids}
\noindent
Mobility and sensory aids are a type of assistive technology that supports a patient by using technology to compensate for their disability~\cite{mezghani}.  Mobility aids focus on a person's physical aspects that pertain to motor skills. Meanwhile, sensory aids provide support towards a patient's sensing and perceptive capabilities. Wearable devices focus on biosensing and the collection of data. Services related to these devices are usually passive in interacting with the patient. On the other hand, aiding devices provide more interactive real-time care for the patient by giving pro-active support for its user.

An example of IoT-based mobility aid is in~\cite{saadeh}. It is a fall predicting and detecting system aiming to assist patients with conditions that cause them to lack coherent motor skills. Patients such as the elderly or those under rehabilitation for trauma are susceptible to weak motor skills. This system aids them by notifying healthcare providers of patients possibly falling or having fallen. With support vector machines, the system can map out the movement tendencies of each patient. As a result, the system can predict and preemptively notify the healthcare service of falling or injury. 
    
\subsection{Smart Devices}
\noindent
These are the devices that have evolved to a more diverse selection of personal devices such as smartphones and other smart accessories~\cite{rpm-smartphone}. A smart device is a more interactive and autonomous sensor since it contains more complex processing capabilities. As a result, assistive technologies have become more convenient for the patient without requiring specific hardware to setup. Unlike the previous two types of devices, smart devices allow the processing of data within the device. Instead of simply collecting data and relying on the server to analyze the results, it can provide a faster diagnosis to the patient. Due to its advanced processing capabilities, it can integrate more sensors and aids to its operations.

An example of a design that uses the automation properties of smart devices is in~\cite{rpm-auto}. They analyze their patient's activity through mobile device sensing paired with cloud computing. With them leaning towards using smart technology to automate their services, they create a system that can cater to each patient in an organized manner and is self-sustaining. Also, they demonstrate the power of these intelligent sensors in creating a more responsive and adaptive system. 
A summary of the different types of assistive technologies in IoT-based healthcare services is shown in Table~\ref{assist-table}.

\begin{table}[t!]
\centering
\normalsize
\begin{tabular}{|l|p{6cm}|p{3.8cm}|}
\hline
\textbf{Types}  & \textbf{Description}& \textbf{Examples} \\ \hline
    \begin{tabular}{l}
    Wearable Device \\ 
    \cite{rpm-iot-ecg, work-low}
    \end{tabular}
    
    & 
    \begin{tabular}{p{5.5cm}}
    Wearable devices that are used to specifically monitor the different physiological metrics of a patient.
    \end{tabular}
    & 
    \begin{tabular}{l}
    - Electrocardiogram \\ - Heart rate sensor \\ - Blood pressure sensor
    \end{tabular}
    \\ \hline
    \begin{tabular}{l}
    Mobility and Sensory Aids \\ 
    \cite{mezghani, saadeh} 
    \end{tabular}
        & 
    \begin{tabular}{p{5.5cm}}
    Medical devices that aid patients who are subject to motor or cognitive disabilities.
    \end{tabular}
    &
    \begin{tabular}{l}
    - Fall sensor \\ - Vision support \\ - Pain sensor  
    \end{tabular}

     \\ \hline
    \begin{tabular}{l}
    Smart Systems \\ 
    \cite{rpm-auto, rpm-smartphone}
    \end{tabular}
    
    &
    \begin{tabular}{p{5.5cm}}
    Smart devices and technology are used to provide interactive aid towards recovering and ailing patients.
    \end{tabular}
    & 
    \begin{tabular}{l}
    - GPS \\ - Assistive robots \\ - Smartphones
    \end{tabular}
    \\ \hline
\end{tabular}
\caption{Examples of assistive technology in IoT-based healthcare.}
\label{assist-table}
\end{table}

\section{Challenges in IoT-Based Healthcare Services}\label{chal}
\noindent 
Incorporating IoT network in healthcare services are beneficial in expanding the ability of medical centers in ensuring the health of their patients. However, this integration comes with a few caveats. The following discusses the IoT-related challenge in patient privacy, data latency, service interactability, device constraints, and scalability in current implementations that use IoT-based healthcare services.

\subsubsection{Patient Privacy}
\begin{figure}[t!]
\centering
\subfloat[Man-in-the-middle.]
{   \includegraphics[width=0.60\columnwidth]{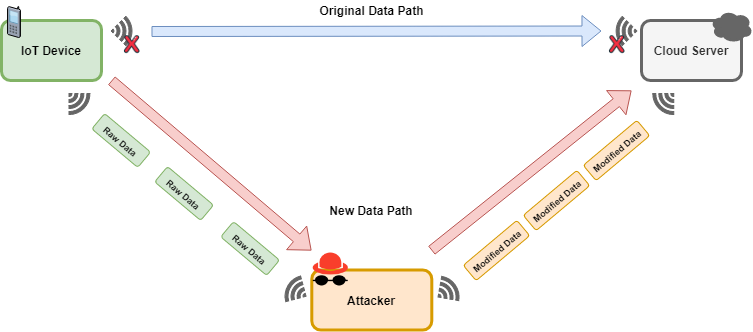}\label{mitm}}\\
\subfloat[Spoofing.]
{    \includegraphics[width=0.70\columnwidth]{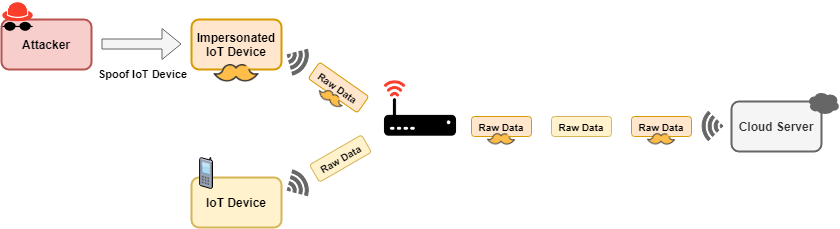}\label{spoof}} \\
\subfloat[Injection.]
{    \includegraphics[width=0.55\columnwidth]{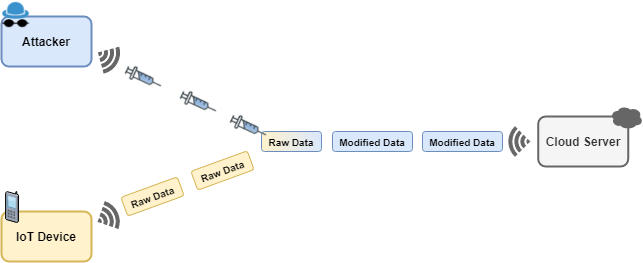}\label{inject}}
\caption{Examples of security attacks caused by a vulnerable data channel.}
\label{attacks}
\end{figure}

The first challenge is about the security of the patient's information. The privacy of the users is important for any network-based online service~\cite{feng-priv}. Even with standard healthcare service, patient confidentiality has always been a strictly valued policy. However, monitoring devices collect the patient's information through the network. This data path adds a transmission layer that is vulnerable to security attacks~\cite{Abbas}. Also, due to the nature of wireless networks, data is being transmitted over long distances. This vulnerability exposes the data to further security attacks. An exposed data channel could lead to patients having their personal information leaked to malicious individuals. The following are examples of security attacks that are caused by a vulnerable data channel~\cite{wiresec-survey}:
\begin{enumerate}[wide, labelwidth=!, labelindent=0pt]
    \item \textit{Man-in-the-middle}. It is when an attacker modifies the transmitted data between the source and its destination. It is carried out by stealthily compromising the data channel and gaining access to the data being sent. This data is then tampered with and sent to its destination without triggering any suspicion from the concerned parties. Attackers can use this to steal and modify patient information before it reaches the medical center. A visual representation of the infiltrating capabilities of this attack is shown in Fig.~\ref{mitm}. 
    
    \item \textit{Spoofing}. It is when an attacker impersonates a compromised user. By gaining access to someone's data, the malicious individual can assume their identity. As a result, the attacker potentially gains access to the user's online accounts and assets. In a healthcare service, an attacker can spoof an account and gain unauthorized access and control on their records and provided service.  A visual representation of the impersonating capabilities of a spoofing attack is shown in Fig.~\ref{spoof}.
    
    \item \textit{Injection}. It is when the attacker inputs wrong or modified information in place of the factual data from the user. Compromised channels are vulnerable to injected fake information that confuses services and misinformation in services. Healthcare centers with vulnerable channels can result in servers receiving the wrong information. As a result, treatments can be delayed and diagnosis can be inaccurate. A visual representation of the tampering capabilities of an injection attack is shown in Fig.~\ref{inject}.
\end{enumerate}

These are some of the potential security concerns that are under the responsibility of the healthcare centers that manage the patient's information. Compromised networks are detrimental to the quality of the service as well as the well-being of the patient~\cite{shuai}. As a result, people are less likely to sign up for these services due to fears of their information being exposed to the network. Without a secure means of maintaining the privacy of data, people will remain skeptical of IoT-based healthcare services. Therefore, in an era where network security is a big concern, these services need to make sure that they focus on fortifying their security to maintain patient data privacy. 

\subsubsection{Data flow}
The next challenge points to the ability of the network to maintain a consistent flow of data. Healthcare services that use IoT-based healthcare services need to receive patient data at a reliable rate to be effective~\cite{work-trans}. As a result, IoT-based healthcare services incorporate wearable devices that observe their patients. These devices collect the patient data and send them through the wireless network. This data transmission is carried out for the duration of patient monitoring. 

Another requirement for the data flow of IoT-based healthcare services is for the data to come in real-time. This allows applications that monitor the patient to be able to effectively check up on the user routinely. With a continuous influx of data from multiple devices, the IoT-based healthcare service could run into issues with server overloading~\cite{rpm-smartphone}. An example of a notable attack that could result in server overloading is a Denial of Service (DoS)~\cite{wiresec-survey}. This attack is when a server is flooded by an influx of data from multiple endpoints. As a result, the service is rendered unable to function as intended. DoS attacks are triggered by a malicious user. However, there are other ways to overload a server. Another means is by reaching the capacity of the network. By having too many endpoints that are transmitting data at the same time, a server will not be able to cope~\cite{fog-cloud}. Without a proper way of regulating this traffic, the service will be unable to function. Therefore, to be able to combat both sources of overloading, a more reliable and adaptive medium for continuous and real-time data transport is needed. 

\subsubsection{Service interactability}
Another challenge is the interactability between the service and the patient within the IoT-based healthcare service. These services lack a proper means of allowing the healthcare service to interact with its patients in real-time~\cite{health-data-int}. Before the incorporation of IoT, the only option for patients was to be physically present within a medical facility to be monitored and diagnosed. Unfortunately, there are situations when the patient is not able to comply. For example, the elderly or disabled have issues with continuous or long travel~\cite{niyato}. As a result, some doctors opt to do house visits to treat and monitor their patients. However, due to the lack of resources and the current system requiring all medical personnel present in the healthcare facility, this is no longer a valid option. Therefore, monitoring devices were introduced to remotely monitor patients without the need for travel. However, as medical conditions became more complex, raw data was no longer enough to diagnose a patient. Patients were still required to schedule in-person checkups for doctors to get a full understanding of their condition. Therefore, a means of being able to interact with the patient remotely will be more beneficial for the convenience of the patient~\cite{miseikis}. By doing so, medical resources and personnel are less hindered by distance and allocation. 

\begin{figure}[t!]
\centering 
\subfloat[Monitoring system with no service interactability.]
{   \includegraphics[width=0.95\columnwidth]{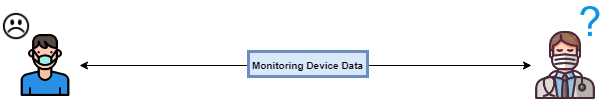}\label{noclue}}\\
\subfloat[Monitoring system with service interactability.]
{    \includegraphics[width=0.95\columnwidth]{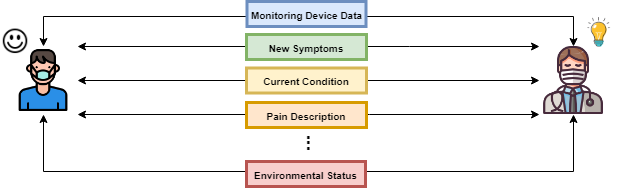}\label{dataclue}}
\caption{Service interactability in monitoring systems.}
\label{serv-int}
\end{figure}

At the same time, as the demand for in-person checkups and treatments is reduced, healthcare resources can be redirected to other patients that have more severe conditions that require full-time in-person monitoring. Without proper interactability, emergencies are harder to handle within IoT-based healthcare services. Without the ability to assess certain real-time situations, healthcare centers will have a difficult time providing support~\cite{yang-pain}. With service interactability, medical centers have a better grasp of the current situation of the patient. Also, data becomes more robust since it is properly defined in the service. With information continuously coming into the server, a means of classifying it before it is analyzed creates a more reliable system for diagnosing anomalies in a user's health. As a result, they have a higher chance of successfully diagnosing any issues or points of concern. A diagram showing the difference between having and missing service interactability in healthcare services is shown in Fig.~\ref{serv-int}. 

\subsubsection{Device Constraints}
The type of IoT devices that are used in the IoT-based healthcare service poses challenges to the application based on its constraints. With the diversity of the IoT technology that is being integrated into the services, each comes with a distinct limitation. Applications such as patient monitoring are usually implemented as long-term, hence, the wearable devices are expected to remain active for the whole time~\cite{work-low}. With devices being required to operate almost indefinitely to maintain constant monitoring of the patient, device efficiency target to sustain sufficient power within the device for long time. The less frequent a device needs to replace a battery or power source, the more it can be effective at carrying out its function within the service. 

Aside from the power capacity of the device, another constraint could also originate from its processor~\cite{rpm-iot-ecg}. With data being transmitted in a continuous and real-time manner, processors need to keep a consistent output of accurate and on-time data. Processors that fail to pass this constraint leads to the disruption of data flow and delay of user diagnosis. It also affects the robustness and reliability of data since the arrival of information to the server is delayed~\cite{deng-delay}.  This results in incoming data that is no longer accurate and unreliable. Another source of inconsistent data from processors could be from the computational resource requirements of modern IoT-based healthcare services~\cite{rpm-iot}. This could be attributed to CPU size and processing technology limitations within the hardware. With newer services incorporating more complex calculations to improve the effectiveness of patient diagnosis and analysis, current wearable devices are required to have some level of processing capability to keep up. However, with the growing concerns in these constraints, incremental solutions to the problem might not be enough to meet the demands for better technology.  

\subsubsection{Scalability}
The last challenge that is discussed in this article is on the scalability of IoT-based healthcare services~\cite{zhang-ubiq}. One of the main reasons for incorporating IoT in healthcare services is to make data and healthcare applications more accessible to its users. As more healthcare centers adapt to the paradigm shift, more devices are added to their network. Also, more patients that transition to using IoT-based healthcare services to manage their health-related records introduce more data to the server. Due to this influx of connecting devices, network traffic, and big data, standard IoT networks are unable to keep up~\cite{mezghani}. This limitation shows the challenge in IoT-based healthcare services in terms of scalability. IoT servers that cannot manage the increasing number of connecting devices run into issues with scaling up. The result is a service that is not able to carry out its intended functionality and purpose. To be able to widen the scope of service, the scalability of the network needs to be improved. Therefore, if healthcare centers intend to increase the reach of their applications, they have to incorporate an IoT-network design that is capable of handling an increasing number of customers.    

\section{ Research Approaches and Opportunities}\label{sol}
\noindent 
The following section highlights how the challenges in patient privacy, data latency, service interactability, device constraints, and scalability can be addressed by improving the IoT medium, as well as some open research opportunities that machine learning can bring in IoT-based applications for healthcare.

\subsubsection{Data Filter}
A critical challenge that IoT-based healthcare services run into is patient privacy. Data is being transmitted wirelessly to a central server from multiple sources. A vulnerable network would lead to the data being compromised, which is detrimental to the privacy of the service's users. A survey in~\cite{Abbas}, focuses on the state-of-the-art approaches related to the e-cloud health system. Their investigation highlights the capabilities of cloud computing as an environment that can relieve infrastructural tasks to a centralized unit. This unit allows the service to minimize costs while making sure that it has control over the data in the network. One of its approaches is the use of precise access control to filter the users that are allowed to access certain levels of data. By incorporating access control to every data transaction and having the permissions centralized, man-in-the-middle attacks are easier to detect with more control over the data being shared between the client and the server. Aside from access control from changing the architecture, some suggest reinforcing the current design. Another take from~\cite{rpm-iot} further dives into the idea of an efficient access control layer for protecting data within the RPM. Their proposed framework uses cryptographic tools as a means to protect and filter the users that attempt to access the data within the server. These tools allow the healthcare service to be protected from attackers that attempt to hack into the service and obtain the patients' data. They use data encryption as a means to hide information as it is being transmitted from the monitoring device to the server. By doing so, only trusted individuals are allowed to view and handle the data. With the creation of a trustworthy system that incorporates access control to protect the data that is being transmitted. Attackers are now not able to effectively impersonate as users since patient identities are more protected by the cryptographic tools used by the framework that was proposed. Lastly, in~\cite{feng-priv} they conducted an investigation on information disclosure being an integral component in patient privacy. By adding a feature that requires the patient to permit to give access to their data, the service adds a convenient means to give users more control over their information. This willingness to disclose information on the medical platform is a crucial step in filtering data within the patient's control. Also, it adds a certain level of intractability within the system to entice the user. By giving a patient more visible control over their information, they will feel safer in being part of the healthcare service. Also, injection attacks are easier to detect since users have control over what information is allowed to enter the server. Any requests for access that is not accounted for and permitted by the user can be easier to detect. This feature allows the network to double-check the source of each data request with the patient and overcome any malicious attacks from unauthorized users. Aside from focusing on what more can the patients do, others propose solutions for improving the architecture of the current system. 

\begin{figure}[t!]
\centering
\subfloat[IoT network without a data filter.]
{   \includegraphics[width=0.95\columnwidth]{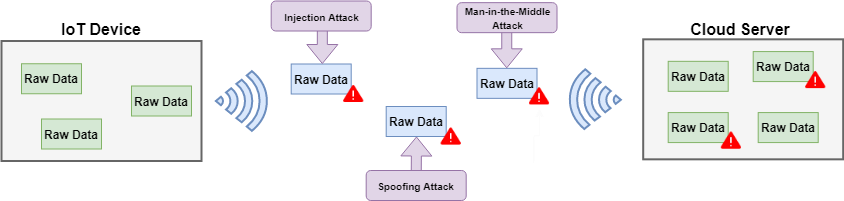}\label{nofilter}}\\
\subfloat[IoT network with a data filter.]
{    \includegraphics[width=0.95\columnwidth]{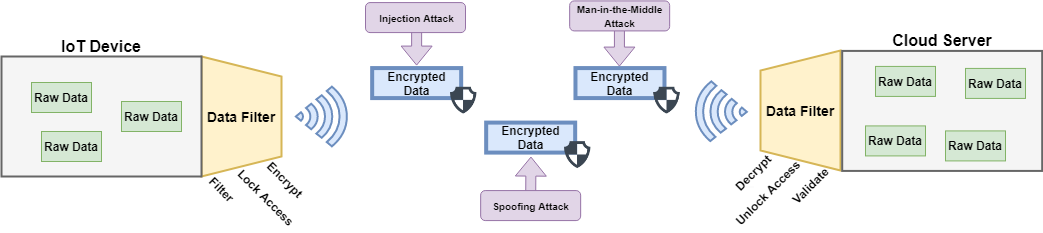}\label{datafilter}}
\caption{Impacts of a data filter on the security of the IoT network.}
\label{filter}
\end{figure}

Overall, these approaches have an underlying theme, which is the creation of a security layer that filters users and data within the server. The main purpose of this data filtering layer is to reinforce the security of the system. Each approach shows its ability to address each type of attack that was discussed in the previous section. Proper data path control for network server aids in minimizing man-in-the-middle attacks. On the other hand, hiding data behind encryption and cryptographic tools allow patients to be protected from spoofing. Lastly, giving more control towards the user to regulate the data that they release to the server also allows the network to check for any malicious behaviour that is out of the ordinary or not permitted by the patient. Thus, this allows the network to be defended from injection attacks. Each cited solution and approach were presented as state-of-the-art means of reinforcing the security of IoT-based healthcare services. This reinforcing is done by preventing potential vulnerabilities via either administrative access control, cryptographic walls, or patient consent. This creates a data filter that will serve as a gate similar to a firewall that will only allow certain types of data to flow through. In turn, data that travel through the network can be better regulated. A visual representation of the impacts of adding a data filter to the network is shown in Fig.~\ref{filter}.

\subsubsection{Data Path Control}
\begin{figure}[t!]
\centering
\subfloat[Cloud network overloaded by processes.]
{   \includegraphics[width=0.7\columnwidth]{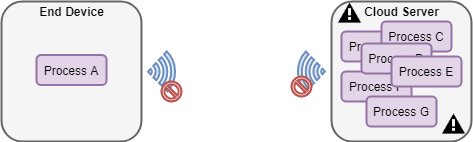}\label{cloud-over}}\\
\subfloat[Fog network with reallocated processes.]
{    \includegraphics[width=0.9\columnwidth]{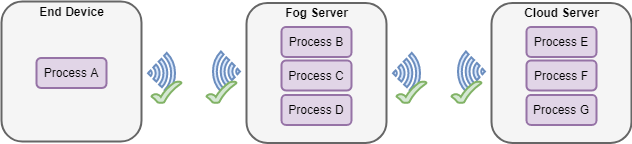}\label{fog-rea}}
\caption{Reallocative properties of adding a fog server to the network.}
\label{reallo}
\end{figure}

The next challenge that was discussed was the concerns in the data latency. With data being continuously transmitted from the wearable devices, the IoT-based healthcare servers can be subject to overloading and disruptions in the flow of data. Any form of delay to the flow of data can cause the quality of the service to diminish. As a result, the response time of these services towards emergencies and diagnosis will become lacking. Al Disi et al. mentions the instability of a cloud-based IoT network in regulating data for online healthcare services~\cite{rpm-iot-ecg}. Instead, they proposed the idea of reallocating certain processes to the IoT gateway. This idea of reallocation is attributed to including an intermediate processing unit between the end device and the cloud. They then incorporate fog technology as a solution to allow the ability to move processes within the network hierarchy. Focusing more on the fog aspect, Verma et al. propose a similar approach~\cite{fog-rpm}. However, their proposal highlights the offloading ability of a fog server in a smart home environment. By taking the distributive capabilities of a fog-based network, resources can be properly reallocated to reduce server strains. Also, fog servers can be used to implement parallelism. It is when exact copies of an entity are used to execute the same task at the same time to increase throughput. As throughput is increased, the tendency of the sever being bottlenecked by the incoming data is lessened. A visual representation of the effects of reallocation to a server overloaded with processes is shown in Fig.~\ref{reallo}. Similar to Verma's approach to using fog computing, Sood et al. propose a Fog-based IoT framework for a decentralized blood pressure monitoring system~\cite{sood}. The fog device allowed them to move some processes closer to the local server for faster detection of emergencies. Instead of waiting for the mail server to respond in case, high blood pressure is detected, the local fog servers are equipped with this capability for faster detection. This feature highlights the decentralized nature of fog-based IoT networks. Decentralization allows a network to give independence to its servers. As a result, local fog servers can function accordingly without the need to consult the main server. This freedom allows processes to be carried out faster, which helps the patient receive treatment on time. Also, decentralization through fog servers can reduce the number of entry points of data to the server~\cite{fog-cloud}. This shift in the network hierarchy can reduce the overall amount of data that is being transmitted directly to the server. Partnered with the data filter that was previously discussed, a means of controlling the data path can help alleviate network strains due to potential overloading.  

\subsubsection{Smart Interactive Systems}
Another challenge that was mentioned was service interactability. IoT-based healthcare services provide convenience to the patient in terms of distance. Also, it makes resources easier to redistribute for healthcare centers. However, to improve the experience for the patients and the help that the service can provide, a more interactive system is needed. Ding et al.~\cite{ding-int} propose a more interactive means of bridging the medical center and the patient. By using different peripherals such as security cameras and smart appliances, more parameters can be added to monitor a patient remotely. As a result, the system can become more adaptive and can cater to a multitude of situations that the modern RPM system can't do. These proposed design choices will allow the healthcare center to monitor the patient closely. It also eliminates the need to require the use of third party communication mediums, since the embedded devices can be integrated into the service. Yang et al.~\cite{yang-pain} proposes an IoT-based system that increases interactability by detecting facial expressions. This model allows the service to develop automatic pain assessment tools when monitoring patients. It shows how behavioural monitoring is also made possible with IoT networks and the peripherals that it can incorporate. Another interactive design is proposed by Miseikis et al.~\cite{miseikis}. They incorporate robotics to provide a therapeutic assistant that can interact with patients. This design allows remote therapy and assistance for patients with limited movement. With its technological capabilities, it can also be used to carry out tasks such as disinfection and elevated boy temperature detection.

These proposed frameworks used interactive mediums to aid the system to connect the patient with the center without concerns of distance, clinical resources, and space. Instead of conducting checkups through phone calls or calling in the patients for physical checkups, these frameworks propose a smart interactive means of administering medical care towards patients using IoT-based healthcare services. Especially for future events, similar to the recent CoViD-19 pandemic, where healthcare centers are being overloaded and people are required to stay indoors. This design creates a smarter healthcare service that allows patients that are not able to travel for checkups the ability to seek medical help. Also, it allows clinics that have capacity and resource issues to be able to take in patients without pushing them into queues and waiting lists. Overall, these designs open the possibility for healthcare centers to expand their IoT-based services, allowing them to treat a larger number of patients.

\subsubsection{Smartphones and Parallel Programming}
The next challenge is from the device constraints found within the IoT devices. Limitations such as power consumption and processor capabilities limit the progress of IoT-based healthcare services. Instead of designing a new wearable device every time an improvement is needed, some designs have moved to incorporate other available technology instead. An example of these technologies is smartphones. Smartphones can be equipped with sensors that allow the collection of health-based metrics. Guo et al.~\cite{guo-biosense} take advantage of the capabilities of smartphones by designing an attachable dongle for biosensing. With the smartphone, power consumption is less of an issue due to its already efficient design. Also, by using ultralow consumption sensors, the overall battery life of the phone is not impacted. Meanwhile, with the use of the smartphones' processor to analyze the collected data, the need to create one for the sensor is no longer needed. This design minimizes the constraints of power consumption and processing power by taking advantage of the already advanced design of the smartphone. Another similar approach is proposed by 

\subsubsection{Smart and Adaptive Environments}
The final challenge that was highlighted in this article was scalability. With the increase of devices that are using IoT-based healthcare services, the growing network becomes harder to manage. To improve the state of the IoT network to handle this challenge, a more adaptive environment is needed. Sun et al.~\cite{sun-scc} talks about the increase in big data sources such as healthcare and safety. They proposed to use smart cities and connected communities to manage this influx of data and devices. By introducing a smart environment where IoT devices can be connected wirelessly, data and processes can be shared effectively. Also, by using the smart capabilities of the environment, the network can be made adaptive to the devices that connect to each implemented healthcare service. Zhang et al.~\cite{zhang-ubiq}, proposes the usage of ubiquitous WSNs to create a more standardized network of sensing devices for the IoT-based healthcare services to use. This allows the incorporation of a wide variety of sensors that together are capable of real-time technical sensing. With the network being able to increase its scope through the incorporation of ubiquitous WSNs, handling the increase of users will be less of a challenge. Partnered with smart environments, regulating the influx of devices can be attainable through its adaptive architecture. Further investment towards an adaptive medium results in the incorporation of automative technologies such as blockchains and deep learning. Kumar et al.~\cite{kumar-block} proposes a design using blockchain technology to improve the management of health records within a healthcare service. With the use of smart contracts, it can automate the management of access control for each device. Partnered with the distributive nature of blockchains, the regulation of resources within the network is made more efficient and effective. Overall, these cited state-of-the-art solutions show the capabilities of technologies such as Smart Environments partnered with deep learning and blockchains in improving the scalability of IoT-based healthcare service. With the use of their automative and adaptive features, each creates an environment that can serve as the main administrator of the devices within the IoT network. 

\subsubsection{Machine Learning Opportunities in Healthcare}
In conjunction with the idea of providing a smarter interactive system, another approach in solving service interactability could be from the incorporation of machine learning (ML) techniques and neural networks. ML is a learning tool used using collected data and establishes trends through training and system modelling. Using these behavioural models, results from similar events can be inferred or predicted. The usage of ML via State Vector Machines (SVM) and an Adaptive-Network-Based Fuzzy Inference System (ANFIS) model to create an IoT-based smart health monitoring and surveillance framework for COVID-19 risk exposure detection is proposed in~\cite{vedaei-ml}. With the use of these analytic ML techniques to train the model, the framework can monitor a registered patient and their physical condition. Also, with the infection data that it has collected and trained with, it can notify them of any risks and direct them towards proper social distancing. Their model claims the ability to regulate the safety of their patients through a smart wearable armband. It also helps those around them to makes sure that the spread of the virus is minimized. This framework shows the potential of ML to model the trends of health-based data and with the use of IoT networks, it can improve the safety and care towards each patient. 

A deep learning approach is proposed in~\cite{amin-cog}. By using 2 Convolutional Neural Network (CNN) models, they were able to detect and classify pathologies within EEG readings. This model services as a cognitive system that is capable of improving the quality and effectiveness of the detecting service for the patient.  Also, with the use of analytic tools such as ML and neural networks, data can be used to refine the overall diagnosing experience within healthcare services.

\section{Conclusion} \label{conc}
\noindent 
Incorporating IoT devices and assistive technologies in healthcare services have several advantages but also poses unique challenges. The issues in patient privacy, data flow, service interactability, device constraints, and scalability are discussed in this article. Related implementations and frameworks were investigated to discover improvements to the IoT network to address these issues. At the same time, the use of ML tools and techniques can improve the effectiveness of regulating the provided healthcare services and their patients. As a result, faster reactions towards risks and emergencies are made possible. Instead of having the cloud server decide if the data warrants any action, the smart wearable devices can now detect anomalies and emergencies earlier by moving the detection layer closer to the patient. Also, it opens to the possibility to use robotics, smart appliances, ML, and neural networks to assist ailing patients as it monitors them as well. Overall, IoT devices provide a window to a more secure, efficient, and interactive system, when they are properly used. Their capabilities are open to more opportunities for research and developing healthcare services to cater to the evolving medical technology.  

\bibliographystyle{IEEEbib}
\bibliography{IEEEabrv,rpmbib} 
\end{document}